\documentclass[aps,pre,showpacs,amsmath,amssymb,twocolumn,10pt,nofootinbib]{revtex4-1}
\usepackage{amsthm}
\usepackage{dsfont}
\usepackage[dvips]{graphicx}

\newcommand{\psib}{\overline{\psi}}
\newcommand{\phib}{{\overline{\phi}}}
\newcommand{\chib}{{\overline{\chi}}}
\newcommand{\Psib}{{\overline{\Psi}}}
\newcommand{\beq}{\begin{equation}}
\newcommand{\eeq}{\end{equation}}
\usepackage{amsmath}
\usepackage{amsfonts}
\usepackage{epsfig}
\usepackage{slashed}
\usepackage{epstopdf}

\begin{document}

\title{Non-abelian gauged NJL models on the lattice}

\author{Simon Catterall, Richard Galvez, Jay Hubisz, Dhagash Mehta, Aarti Veernala}
\affiliation{Department of Physics, Syracuse University, Syracuse, NY 13244, USA}

\date{\today}

\begin{abstract}
We use Monte Carlo simulation to probe the phase structure of
a $SU(2)$ gauge theory containing $N_f$ Dirac fermion flavors transforming in the fundamental representation of the group and interacting through an additional four fermion term.  Pairs of physical flavors are
implemented using the two tastes present in a reduced staggered fermion
formulation of the theory. The resultant lattice theory
is invariant under a set of shift symmetries which correspond to a discrete subgroup of the
continuum chiral-flavor symmetry. The pseudoreal character of the representation guarantees that the theory
has no sign problem.
For the case of $N_f=4$ we observe a
crossover in the behavior of the chiral condensate for strong
four fermi coupling associated with the
generation of a dynamical
mass for the fermions. At weak gauge coupling this crossover is consistent with the usual continuous phase
transition seen in the pure (ungauged) NJL model. However, if the gauge coupling is strong enough to cause confinement
we observe a much more rapid crossover in the chiral condensate
consistent with a first order phase transition\end{abstract}


\maketitle

\section{Introduction}
Elucidating the nature of the electroweak symmetry
breaking sector of the Standard Model (SM) is the main goal of the Large Hadron Collider currently running at CERN. It
is widely believed that the simplest scenario involving a single scalar Higgs field is untenable due to the fine tuning and triviality problems which arise
in scalar field theories.  One natural solution to these problems can be found by assuming that the Higgs sector
in the Standard Model arises as an effective field theory describing the dynamics of a composite field arising
from strongly bound fermion-antifermion pairs.

One class of models that have been proposed which exhibit these features are technicolor theories (TC) \cite{TC-intro, TC-intro2}
in which a new non-abelian
gauge interaction causes the condensation at low energy of  fermion bound states which are presumed to carry electroweak quantum numbers.  These condensates break the electroweak gauge group, giving mass to the $W$ and $Z$ bosons.  The realization that theories of this
type utilizing fermions in two index representations of the gauge group
may offer an explanation of dynamical symmetry breaking which is {\it not} at variance with
electroweak precision measurements \cite{walking-francesco1} has led to numerous recent lattice
studies - see the conference reviews \cite{Fleming:2008gy, Pallante:2009hu, DelDebbio:2010zz,Fodor:2010zz} and references therein.

However, to obtain fermion masses in these scenarios requires additional model building, as in extended technicolor models~\cite{ETC-1,ETC-2,schrock2,schrock3} and models of top-condensation~\cite{Miransky:1988xi,Miransky:1989ds,Bardeen:1989ds,Marciano:1989mj}. In the latter
models four-fermion interactions drive the formation and condensation of a scalar top--anti-top bound state which plays
the role of the Higgs at low energies.

Our motivation in this paper is to study how
the inclusion of such four fermion interactions may influence the
phase structure and low energy behavior of non-abelian gauge theories in general.
Specifically we have examined a model with both gauge interactions and a chirally invariant
four fermi interaction - a model known in the literature as the gauged NJL model \cite{DSB-yamawaki}. The original
NJL model \cite{NJL-org} without gauge interactions is known to exhibit spontaneous
breaking of chiral symmetry for large four fermi coupling.   These models have been studied extensively on the lattice~\cite{Hands:1997uf,Hands:1997xv}.   In the vicinity of the phase transition between chirally symmetric and broken phases,
the theory is thought to be renormalizable and to correspond to an
elementary scalar field theory coupled to fermions \cite{annakuti}.  As such,  the continuum
limit is believed to be governed by the usual IR attractive
gaussian fixed point characteristic of scalar field theory\footnote{The authors wish to thank Julius  Kuti and Anna Hasenfratz for important discussions on these issues}.The abelian gauged NJL model has been studied on the lattice as well~\cite{Kim:2001am}, in which numerical evidence for the triviality of QED was presented.

The focus of the current work is to explore the phase diagram when fermions are charged under a non-abelian gauge group.  Indeed, arguments have been given in the continuum that
the gauged NJL model may exhibit different critical behavior at the boundary
between the symmetric and broken phases \footnote{Notice that the appearance of a true
phase transition in the gauged NJL models depends on the approximation that we can neglect the
running of the gauge coupling} corresponding to the appearance of a line
of new fixed points associated with a mass anomalous dimension varying in the
range
 $1 < \gamma_{\mu} < 2$ \cite{DSB-yamawaki,walking-francesco2}. The evidence for this behavior
derives from calculations utilizing the ladder approximation in Landau gauge to the Schwinger-Dyson equations.  A primary goal of the current study was to use lattice simulation to check the validity of
these conclusions and specifically to search for qualitatively new critical behavior in the
gauged model
as compared to the pure NJL theory. While we will present results that indicate that
the phase structure of the gauged NJL model is indeed different from pure NJL, we shall
argue that our results are \emph{not} consistent
with the presence of any new fixed points in the theory.

To facilitate this study we have chosen to employ a reduced staggered fermion lattice formalism. This has the advantage of allowing us
to incorporate as few as two continuum flavors of Dirac fermion and, as we will show in Section~\ref{latticemodel}, allows us to build
in  lattice four fermi terms which are invariant under a discrete subgroup of
the continuum chiral symmetries~\cite{redstag-Smit-1,redstag-Smit-2}. The presence of four fermi interactions has an additional attractive feature - it allows
us to study the lattice theory with exactly zero fermion mass~\cite{Kogut:1996mj}. Thus the observation of a non zero condensate
corresponds, in the infinite
volume limit, to a spontaneous breaking of lattice
chiral symmetry and the dynamical generation of quark masses.
This  discrete symmetry breaking should correspond in the continuum limit to a breaking of
the usual continuous chiral-flavor symmetry. 
The price one
pays for this simplicity is that the  lattice
fermion operator possesses small eigenvalues (at least for small four fermi coupling) and it has only been possible to study
modest lattice volumes using a GPU accelerated code.  Nevertheless the results show no strong volume dependence and
should give a robust indication of the phase structure of the theory in the infinite volume limit.

In the work reported here we have concentrated on the four flavor theory
corresponding to
two copies of the basic Dirac doublet used in the lattice construction.
The four flavor theory is expected to be chirally
broken and confining at zero four fermi coupling. Understanding the effects of the four fermion term in this
theory can then serve as a benchmark for future studies of theories which, for zero
four fermi coupling, lie near or inside the conformal window. In the latter
case the addition of
a four fermion term will break conformal invariance but in principle that breaking may be made arbitrarily small
by tuning the four fermi coupling. It is entirely possible that the phase diagrams of such conformal or walking
theories in the presence of four fermi terms may exhibit very different features than those seen for
a confining gauge theory.

In the next section we write down the continuum theory we are studying and explain how to rewrite it in a more convenient
{\it twisted basis} in which the two usual Dirac spinors of the theory are replaced by a single matrix valued fermion field.
This is the same transformation that lies at the heart of recent efforts to construct lattice theories with
exact supersymmetry \cite{exact-latt-susy} and corresponds also to the spin-taste representation of staggered fermions \cite{redstag-Smit-1}.
We then show how to discretize this twisted two fermion theory to arrive at a reduced staggered fermion lattice theory which incorporates the
Yukawa interactions needed to generate the four fermi terms \cite{redstag-Smit-2}. We then describe the exact symmetries of the lattice
action relating them to the chiral-flavor symmetries of the continuum theory. The pseudoreal character of the fundamental
representation of the
$SU(2)$ group allows us to avoid a potential sign problem after integration over the
fermions.

We then go on to describe our numerical results
on the phase diagram for the four flavor theory.
We have simulated the model by sweeping in the four fermi coupling for a fixed gauge coupling.
A series of these gauge couplings were examined which span the range from confined to deconfined regimes of
the gauge theory in the absence of four fermion terms.
We show that the chiral
phase transition expected in the simple NJL model changes character in the gauged model;
strictly speaking the gauge model (at least for four flavors) already breaks
chiral symmetry spontaneously even for zero four fermi coupling so that no
true transition is present. Nevertheless we observe a very rapid crossover behavior
for strong four fermi coupling and recover evidence for would be Goldstone bosons above the
crossover region.  We see no evidence for the existence of new UV fixed points in the theory.

\section{Continuum gauged NJL model}
\label{sec:Continuum}

We will consider a model which consists of $N_f/2$ doublets of gauged massless Dirac fermions in the
fundamental representation of an $SU(2)$ gauge group and
incorporating an $SU(2)_L\times SU(2)_R$ chirally invariant four fermi interaction.
The action for a single doublet takes the form
\begin{eqnarray}
S &=& \int d^4x\; \psib ( i \slashed{\partial} -  \slashed{A}) \psi - \frac{G^2}{2N_f} [ (\bar{\psi} \psi)^{2} + (\bar{\psi} i \gamma_{5} \tau^{a} \psi )^{2} ]  \nonumber \\
&-& \frac{1}{2g^2} Tr [F_{\mu \nu} F^{\mu \nu}] ,
\label{eq:etcnjlaction}
\end{eqnarray}
where G is the four-fermi coupling, $g$ the usual gauge coupling
and $\tau^{a},a=1\ldots 3$ are the generators of the $SU(2)$ flavour group.

\begin{figure}
\begin{center}
\includegraphics[width=0.75\linewidth]{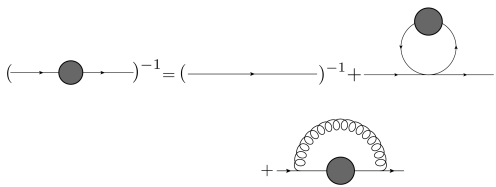}
\caption{Diagramatic truncated Schwinger-Dyson equation for the fermion self-energy}
\label{fig:SDdiagrams}
\end{center}
\end{figure}

This theory has been explored in the continuum using approximations to the Schwinger-Dyson equations in which sub-classes of planar loop diagrams are re-summed.  This ``ladder" approximation neglects the running of the four-fermion interaction, and treats the running of the gauge coupling in only a heuristic way, implementing the momentum dependence of the non-abelian gauge coupling by hand.  In this approximation, the Schwinger-Dyson equation for the fermion two point function is \begin{widetext}\begin{equation}
S_F^{-1} [\displaystyle{\not}p] = (S_F^{(0)} [\displaystyle{\not}p])^{-1} + i G^2\int \frac{d^4k}{(2 \pi)^4} \text{Tr}~S_F [\displaystyle{\not}k] -  4 \pi C_2 (F) \int \frac{d^4k}{(2 \pi)^4} \alpha [p-k] \gamma^\mu S_F [\displaystyle{\not}k] \gamma^\nu D_{\mu \nu} [p-k] ,
\end{equation} \end{widetext}
where the gauge boson propagator is taken to be in Landau gauge, such that this self-energy term is finite without taking into account vertex corrections.  This equation is expressed diagrammatically in Figure~\ref{fig:SDdiagrams}.  A typical ansatz for the functional form of the gauge coupling is \begin{equation}
\alpha[p^2] = \left\{ \begin{array}{cl} \frac{\alpha[\Lambda^2]}{1- \alpha[\Lambda^2] \frac{b}{8 \pi}  \log \frac{p^2}{\Lambda^2}} & p^2 > \Lambda_c \\ \alpha (\Lambda_c^2) & p^2 \le \Lambda_c \end{array} \right.,
\end{equation}
where $b$ is the $\beta$-function coefficient, and $\Lambda_c$ is the confinement scale for the gauge theory.

Approximate analytic numerical solutions for the fermion self-energy were studied in~\cite{Takeuchi:1989qa}.  It was argued that in asymptotically free theories which themselves confine and generate a chiral condensate, the second order NJL phase transition in the ungauged NJL case morphs with increasing gauge coupling into a cross-over phenomenon where the chiral condensate is dramatically enhanced at a critical value for the four fermi coupling.  Our analysis constitutes a non-perturbative exploration of this crossover phenomenon in the two dimensional parameter space of the bare gauge and four fermi
couplings.

To implement the theory in Eq.~\ref{eq:etcnjlaction} on the lattice, it is convenient to reparametrize the four fermi term in the continuum action via the use of scalar auxiliary fields. Specifically the fermion interaction term is replaced by Yukawa and scalar mass terms
\begin{equation}
S_\text{aux} =\int d^4x\; \frac{G}{\sqrt{N_f}}\left(\bar{\psi}\psi\phi_4+\bar{\psi}i\gamma_5\tau^a\psi\phi_a\right)+\frac{1}{2}\left(\phi_4^2+\phi_a^2\right).
\end{equation}
The action is now quadratic in the fermions and that portion of the path integral can be performed analytically.  This yields the usual fermion determinant as a function of the scalar field configurations which are then numerically integrated over.

It is convenient when we come to discretization to rewrite the fermionic sector of this theory in terms of a new set of
matrix valued fields. To see how these arise consider first a system of four Dirac fermions
$\psi^{i}_{\alpha}$ where $i=1\ldots 4$ is a flavor
index and $\alpha$ a spinor index (initially consider a model without Yukawa
interactions).
If we denote the matrix implementing the usual space-time rotations  by $R_{\alpha\beta}$
and the corresponding one for flavor rotations by $F^{ij}$ then these fermions transform as
\begin{equation}
\psi^i_\alpha=R_{ij} F^{\alpha\beta}\ \psi^j_\beta .
\label{eq:symmetries-1}
\end{equation} Using only lower indices this can be trivially rewritten as
\begin{equation}
\psi_{i\alpha}=R_{ij} \ \psi_{j\beta} \ F^{T}_{\beta\alpha} .
\label{eq:symmetries-2}
\end{equation} Thus under the diagonal subgroup corresponding to equal rotations in flavor and space, $R=F$,
one can treat the fermions as matrix valued fields, $\Psi$.

In this formalism there is a natural way to reduce the number of degrees of freedom from four to two; introduce the projected
matrices
\begin{equation} \Psi\to \frac{1}{2}\left(\Psi-\gamma_5\Psi\gamma_5\right),\qquad \Psib\to\left(\Psib+\gamma_5 \Psib \gamma_5\right) \end{equation}
More explicitly in a chiral basis this implies that the matrix fields take the block matrix form
\begin{equation} \Psi=\left(\begin{array}{cc} 0&\psi_R\\ \psi_L&0 \end{array}\right),\qquad \Psib=\left(\begin{array}{cc}\psib_L&0\\ 0&\psib_R\end{array}\right). \end{equation} Note that while
$\Psi$ is a $4\times 4$ matrix field the
fields $\psi_{R}$ and $\psi_{L}$ are just $2\times 2$ matrix fields each of which can be thought
of as corresponding to 2 flavors of Weyl fermion. This can be confirmed by
computing the kinetic term which now reads
\begin{equation}
\int {\rm Tr}(\Psib \gamma_\mu\partial_\mu \Psi)=\psib_L\sigma_\mu\partial_\mu\psi_L+
\psib_R\overline{\sigma_\mu}\partial_\mu \psi_R
\label{eq:cont-action-kin}\end{equation}
where $\sigma_\mu=\left(\sigma_i,iI\right)$.
Furthermore, Yukawa type interactions of the form $\psib_L\phi\psi_R+\psib_R\phi^{\dagger}\psi_L$
can also be written in (projected) matrix form as
\begin{equation}
{\rm Tr} \left(\Psib\Psi\Phi\right) ,
\label{yukawacontinuum}
\end{equation} where
\begin{equation}
\Phi=\left(\begin{array}{cc}
0&\phi\\
\phi^\dagger &0\end{array}\right)=\phi_\mu\gamma_\mu ,
\end{equation} with the $2\times 2$ matrix $\phi=\phi_4 I+i\phi_i\tau_i$. These Yukawa interactions
are chirally invariant if the scalar field $\phi_\mu$ transforms appropriately.
In the end we can use these Yukawa terms to build four fermi interactions by adding a quadratic
term for the scalar field of the form $\frac{1}{2}\phi^2_\mu$ and subsequently integrating out $\phi_\mu$.

\section{Discretization on a lattice}
\label{latticemodel}

The reason that we have recast the continuum theory in this language of matrix twisted
fields is that it admits a simple transcription to the lattice where it becomes the
well known reduced staggered formulation of lattice fermions.

We start with the matrix fields $\Psi$ and $\Psib$ introduced in the last section, for the moment
considering the unprojected matrices.  We then expand these matrices on a basis corresponding to
products of gamma matrices and associate these products with staggered fields $\chi$, $\chib$.
\begin{eqnarray}
\Psi(x)&=&\frac{1}{8}\sum_b \gamma^{x+b}\chi(x+b) ,\\
\Psib(x)&=&\frac{1}{8}\sum_b (\gamma^{x+b})^\dagger\chib(x+b) ,
\label{matrixfields}
\end{eqnarray}
where $\gamma^{x+b}=\prod_{i=1}^4 \gamma_i^{x_i+b_i}$ and the sums correspond to the
vertices in an elementary hypercube associated with lattice site $x$ as the components vary $b_i=0,1$ \footnote{If the fields $\chi(x+b)$ are
replaced with $\chi_b(x)$ where the non-zero components of $b$ denote tensor indices for link
fermions running from $x\to x+b$ we recover the recent lattice constructions based on
twisted supersymmetry}\cite{redstag-Smit-2,redstag-Sharatchandra}.
It is easy to see that the projected matrix fields introduced in the continuum construction
then merely correspond to restricting the staggered fields $\chi$ and $\chib$
to odd and even lattice sites respectively via \beq \chi (x) \rightarrow \frac{1}{2} [1- \epsilon(x)] \  \chi (x) \ \ , \ \ \chib (x) \rightarrow \frac{1}{2} [1+ \epsilon(x)]  \ \chib (x), \eeq where, $\epsilon(x) = (-1)^{x_{1}+x_{2}+x_{3}+x_{4}}$.
Furthermore since $\chi$ and $\chib$ now live in different sites on the lattice we refine
$\chib\to\chi$ and consider only a single
staggered field $\chi$.
This restriction of the single component fields $\chi$ and $\chib$ reduces the number of degrees of freedom by a factor of two so the continuum limit of this lattice theory contains two Dirac fermion flavors.
The free reduced staggered kinetic action can therefore be recast as
\begin{eqnarray}
S_{\rm kin}&=&\frac{1}{64}\sum_{x,\mu}\frac{1}{2}{\rm Tr}\left[\Psib(x)\gamma_\mu(\Psi(x+\mu)-\Psi(x-\mu))\right]\\
&=&\frac{1}{64}\sum_{x,\mu,b,b^\prime}\chi(x+b)\chi(x+\mu+b^\prime) \nonumber
\\ & & \times \; {\rm Tr}\left((\gamma^{x+b})^\dagger\gamma_\mu\gamma^{x+b^\prime+\mu}\right)  \nonumber \\
&=&\sum_{x,\mu}\eta_\mu(x)\chi(x)\chi(x+\mu)
\end{eqnarray}
Here, we have substituted the matrix expressions given in Eq.~(\ref{matrixfields}) into
the free Dirac action having replaced the continuum derivative with a
symmetric difference operator and evaluated the
trace as $4\delta_{b,b^\prime+\mu}\eta_\mu(x)$ where
$\eta_\mu(x)$ is the usual staggered quark phase given by \vspace{29pt}
\beq  \eta_{\mu}(x) = (-1)^{\sum_{1}^{\mu - 1} x_{\mu}} . \eeq Gauging
the reduced staggered theory we obtain \cite{redstag-Smit-1}
 \beq S_{kin} = - \sum_{x,\mu} \ \frac{1}{2} \eta_{\mu}(x) [ \chib^{T}(x) \mathcal{U}_{\mu}(x) \chi(x + a_{\mu}) ] \label{Skin-latt} \eeq where
  \beq \mathcal{U}_{\mu} (x) = \frac{1}{2} [1+ \epsilon(x)] \; U_{\mu}(x) + \frac{1}{2} [1- \epsilon(x)] \; U_{\mu}^{*}(x) \label{mathcalU}. \eeq
Finally, the Yukawa interactions from equation (\ref{yukawacontinuum}) on the lattice take the form:
\begin{eqnarray}
S_{\rm Yuk}&=&{\rm Tr}\left(\Psib(x)\Psi(x)\Phi(x)\right)\\
&=&\sum_{x,b,b^\prime}\chib(x+b)\chi(x+b^\prime)\phi_\mu(x) \nonumber \\
& & \times \; {\rm Tr}\left(\gamma^{x+b})^\dagger\gamma_\mu\gamma^{x+b^\prime}\right) \nonumber \\
&=&\sum_{x,\mu}\chi(x)\chi(x+\mu)\phib_\mu(x)\epsilon(x)\xi_\mu(x) ,
\label{yukawa-latt}
\end{eqnarray}
where the trace evaluation now leads to $4\delta_{b,b^{'}\pm\mu}\xi_\mu(x)$ with the
phase $\xi_\mu(x)=(-1)^{\sum_{i=\mu+1}^4x_i}$ and
\beq
\phib_\mu(x)=\frac{1}{16}\sum_b\phi_\mu(x-b) . \eeq
Notice that if we assign the scalar to the dual lattice this latter expression simply represents the average of
the scalar field over the dual hypercube associated with a given lattice site.
Combining Eqs.~(\ref{Skin-latt}) and (\ref{yukawa-latt}), the gauged massless action
including Yukawa interactions can be written in
terms of a reduced staggered field as
\beq
S= \sum_{x,\mu} \ \chi^{T}(x) \ \mathcal{U}_{\mu}(x) \ \chi(x + a_{\mu}) \ [\eta_{\mu}(x) +G\; \phib_\mu(x) \,\epsilon(x) \, \xi_\mu(x)] .
\label{finalS-latt} \eeq The two staggered tastes become the two physical quark flavors in the
continuum limit and as we will see the lattice action possesses additional discrete symmetries
which form a subgroup of the continuum chiral-flavor symmetries.

\section{Symmetries of the lattice theory}

Clearly the theory is invariant under the $U(1)$
symmetry $\chi(x)\to e^{i\alpha\epsilon(x)}\chi(x)$ which is to be interpreted as the $U(1)$ symmetry corresponding to
fermion number.
More interestingly it is also invariant under
certain shift symmetries given by 
\begin{eqnarray}
\chi(x)&\to&\xi_\rho(x)\chi(x+\rho) , \\
U_\mu(x)&\to&U_\mu^{*}(x+\rho) , \\
\phi_\mu(x)&\to&(-1)^{\delta_{\mu\rho}}\phi_\mu(x+\rho) .
\end{eqnarray}
The transformed action is given by
\begin{widetext}\beq
S=\sum_{x,\mu}\xi_\rho(x)\chi(x+\rho) \ \mathcal{U}_\mu(x+\rho) \xi_\rho(x+\mu) \chi(x+\mu+\rho)
\eta_\mu(x+\rho)
\left(1+G \phib(x+\rho)(-1)^{\delta_{\mu\rho}}\right), \eeq \end{widetext}
where we have used the result $\xi_\mu(x)\epsilon(x)=(-1)^{x_\mu}\eta_\mu(x)$ and noted that
any multiplicative phase change in $\phib(x)$ associated with the shift symmetry
is automatically cancelled by the corresponding shift in
the factor $(-1)^{x_\mu}$.
Therefore, shifting the summation vector $x \to x+\rho$, using \beq \epsilon(x) \rightarrow -\epsilon(x+\rho) \eeq  and assuming periodic boundary conditions, the transformed
action can then be rewritten
\beq
S=\sum_{x,\mu} \chi(x)^{T} \ \mathcal{U}_\mu(x) \chi(x+\mu) \left[\xi_\rho(\mu)\eta_\mu(\rho)\right]
\left(1+G \phib(x+\rho)(-1)^{\delta_{\mu\rho}}\right),
\eeq where we have used \begin{eqnarray} \xi_{\rho}(x) \xi_{\rho}(x + \mu) = \xi_{\rho}(\mu) \\
\eta_{\mu}(x+\rho) = \eta_{\mu}(x) \eta_{\mu}(\rho). \end{eqnarray}
It is then not hard to see that the phase factor in square brackets is always unity and the
hence the action is invariant under the original shift symmetry. These shift symmetries
correspond to a {\it discrete} subgroup of
the continuum axial flavor transformations which act on the matrix field $\Psi$ according to
\beq \Psi\to \gamma_5\Psi\gamma_\rho\eeq
Notice that no single site mass term is allowed in this model.

\section{Numerical results}

We have used the RHMC algorithm to simulate the lattice theory  with
a standard Wilson gauge action being employed for the gauge fields. Upon integration over
the basic fermion doublet we obtain a Pfaffian ${\rm Pf(M(U))}$ depending on the gauge field \footnote{Note that the fermion operator appearing in eqn.~\ref{finalS-latt} is antisymmetric}.
The required pseudofermion weight for $N_f$ flavors is then
${\rm Pf}(M)^{N_f/2}$. The pseudoreal character of
$SU(2)$ allows us to show that the
Pfaffian is purely real \footnote{In practice we observe that the Pfaffian is in fact not only real
but also always positive definite so multiples of two
flavors should be possible too.} and so we are guaranteed to have no sign problem if
we use multiples of four flavors corresponding to a
pseudofermion operator of the form  $(M^\dagger M)^{-{\frac{N_f}{8}}}$. The results in this
paper are devoted to the case $N_f=4$.
We have utilized
a variety of lattice sizes: $4^4$, $6^4$, $8^4$ and
$8^3\times 16$ and a range of gauge couplings $1.8< \beta \equiv 4/g^2< 10.0$.
\begin{figure}[htb]
\begin{center}
\includegraphics[height=70mm]{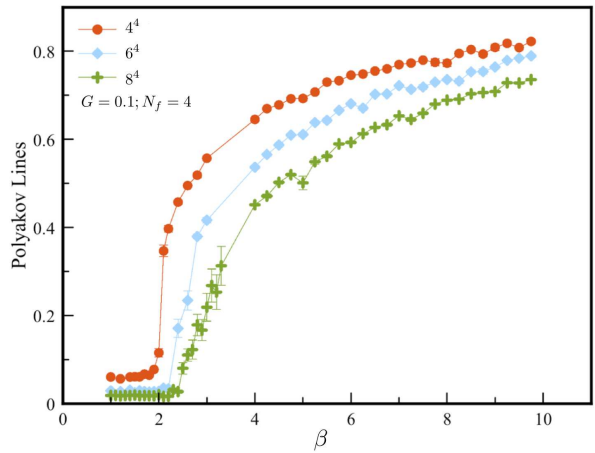}
\caption{Polyakov loop vs $\beta$ at $G=0.1$ for four flavours}
\label{poly-L468}
\end{center}
\end{figure}
To determine where the pure gauge theory is strongly coupled and confining we
have examined the average Polyakov line
as $\beta$ varies holding
the four fermi coupling fixed at $G=0.1$.
This is shown in figure ~\ref{poly-L468}.
We see a strong crossover between a confining regime for small $\beta$
to a deconfined regime at large $\beta$. The crossover coupling is volume dependent and takes
the value of $\beta_c\sim 2.4$  for lattices of size $L=8$.
For $\beta<1.8$ the plaquette
drops below 0.5 which we take as indicative of the presence of strong lattice spacing artifacts and so
we have confined our simulations to larger values of $\beta$.
We have set the fermion mass to zero in all of our work so that our lattice
action possesses the series of exact chiral symmetries discussed earlier.

One of the primary observables used in this
study is the chiral condensate which
is computed from the gauge invariant one link mass operator
\beq
\chi (x)\left({\mathcal U}_\mu(x)\chi (x+e_\mu)+{\mathcal U}^\dagger_\mu(x-e_\mu)\chi(x-e_\mu)\right) \epsilon (x)\xi_\mu(x)\eeq
Because of the absence of spontaneous
symmetry breaking in finite volume we measure the absolute value of this operator. In a chirally broken phase we
expect this to approach a constant as the lattice volume is sent to infinity. Conversely if chiral symmetry is restored
this observable will approach zero in the same limit. In all our runs we observe that the the only component of the auxiliary field to
develop a vacuum expectation value corresponds to the Dirac mass term represented
by the component $\mu=4$. This is consistent with the usual conjecture that the chiral
symmetries break to the maximal subgroup.

\begin{figure}
\begin{center}
\includegraphics[height=70mm]{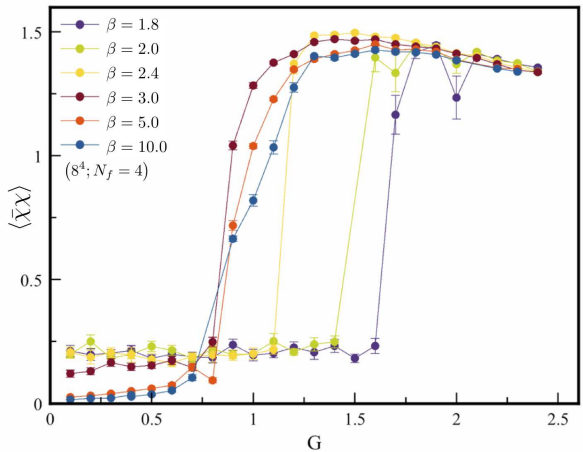}
\caption{$\langle\chib\chi\rangle$ vs $G$ for varying $\beta$ for the $8^4$ lattice with $N_f = 4$.}
\label{oct12-psibpsi-L8-N4-all-beta}
\end{center}
\end{figure}
\begin{figure}
\begin{center}
\includegraphics[height=70mm]{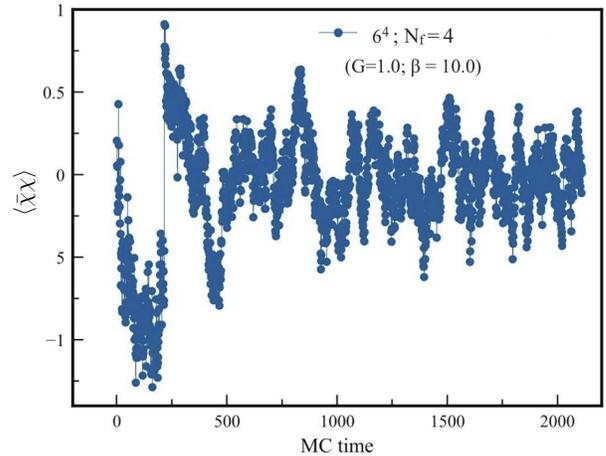}
\caption{$\langle\chib\chi\rangle$ vs Monte Carlo time, t, for $\beta=10.0$ at $G = 1.0$ for the $6^4$ lattice with $N_f = 4$.  Note that here we do not take the absolute value.  In this case, $G = 1.0 \equiv G_{cr}$ is the point at which the transition occurs}
\label{oct28-npsibpsi-L6N4-beta10-G1}
\end{center}
\end{figure}
\begin{figure}
\begin{center}
\includegraphics[height=70mm]{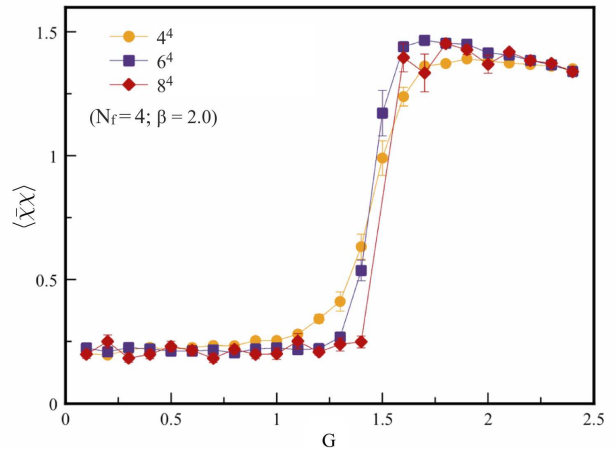}
\caption{$\langle\chib\chi\rangle$ vs $G$ at $\beta=2.0$ for lattices $4^4$, $6^4$ and $8^4$ with $N_f = 4$.}
\label{oct12-psibpsi-all-L-N4}
\end{center}
\end{figure}
\begin{figure}
\begin{center}
\includegraphics[height=70mm]{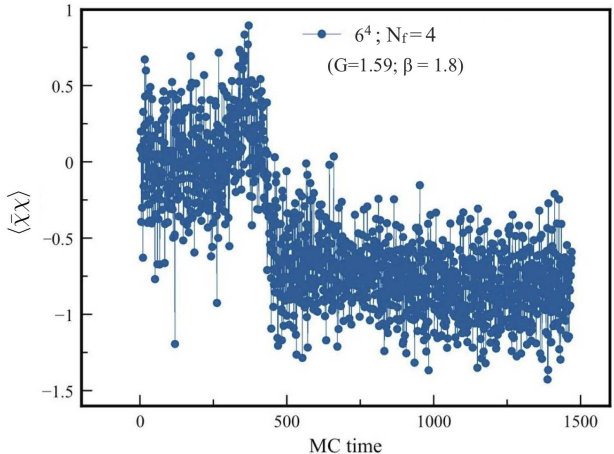}
\caption{$\langle\chib\chi\rangle$ vs Monte Carlo time, t, for $\beta=1.8$ at $G= 1.59$ for the $6^4$ lattice with $N_f = 4$. Note that here we do not take the absolute value.  $G = 1.59 \equiv G_{cr}$ is the point at which the transition occurs.}
\label{oct28-npsibpsi-L6N4-beta1p8-G1pt59}
\end{center}
\end{figure}

In Figure \ref{oct12-psibpsi-L8-N4-all-beta} we 
show a plot of the absolute value of the condensate at a variety of gauge couplings
$\beta$ on $8^4$ lattices.  Notice the rather smooth transition between symmetric and broken phases around
$G\sim 0.9$ for $\beta = 10$. This is consistent with earlier work using sixteen flavors of naive fermion reported
in \cite{annakuti} which identified a line of second order phase transitions in this region of
parameter space. It also agrees with the behavior seen in previous simulations using conventional staggered quarks \cite{Hands:1997uf}.
The second order nature of this transition, for large $\beta$ values, can be confirmed by examining the Monte Carlo time series for the condensate
close to the transition as shown in Figure \ref{oct28-npsibpsi-L6N4-beta10-G1}. Large fluctuations are observed but there
is no sign of metastability or a two state signal in the Monte Carlo evolution.
This behavior should be contrasted with the behavior of the condensate for strong gauge coupling $\beta\le 2.4$. Here a very
sharp transition can be seen reminiscent of a first order phase transition. In Figure~\ref{oct12-psibpsi-all-L-N4} we highlight this
by showing a plot of the condensate versus four fermi coupling at the single gauge coupling $\beta=2.0$ for a range of
different lattice sizes. 
The chiral condensate is now non-zero even for small four fermi coupling and shows no
strong dependence on the volume consistent with spontaneous chiral symmetry breaking in the pure gauge
theory. However, it jumps abruptly to much
larger values when the four fermi coupling exceeds some critical value.
This crossover or transition
is markedly discontinuous in character - reminiscent of a first order phase transition. Indeed,
while the position of the phase transition is only weakly volume dependent it appears
to get sharper with increasing volume.
To try to see whether the jump is indeed first order we have once again examined the Monte Carlo
time series for the condensate close to the jump - the results are shown in Figure~\ref{oct28-npsibpsi-L6N4-beta1p8-G1pt59} for
a lattice with $L=6$ at $\beta=1.8$.
Clearly the system suffers from extremely long relaxation times close to the transition region - only finding the correct
ground state after hundreds of Monte Carlo sweeps. However,
we have not observed a tunneling between two competing
minima as one would expect of a true first order transition and so it is hard to state with certainty that the transition is indeed first order.

What seems clear is that the second order transition seen in the
pure NJL model is no longer present when the gauge coupling is strong.
In the next section we will argue that this is to be expected -- in the gauged
model one can no longer send the fermion mass to zero by adjusting the four fermi coupling since it receives
a contribution from gauge mediated chiral symmetry breaking. Indeed the measured one link chiral condensate operator is not an order parameter
for such a transition since we observe it to be non-zero for all $G$. Notice however that  we see no sign
that this condensate depends on the gauge coupling $\beta$ in the confining regime at small $G$. This is qualitatively different from the behavior  of regular staggered
quarks and we attribute it to the fact that the reduced formalism does not allow for a single site mass term or an
exact {\it continuous} chiral symmetry. Thus the spontaneous breaking of the
residual discrete lattice chiral symmetry by gauge interactions will
not be signaled by a light Goldstone pion and the measured condensate will receive contributions only from massive states.
The transition we observe  is probably best thought
of as a crossover phenomenon corresponding to the sudden onset of a new mechanism for dynamical mass generation due to the strong four fermi
interactions.

\begin{figure}[h]
\begin{center}
\includegraphics[height=70mm]{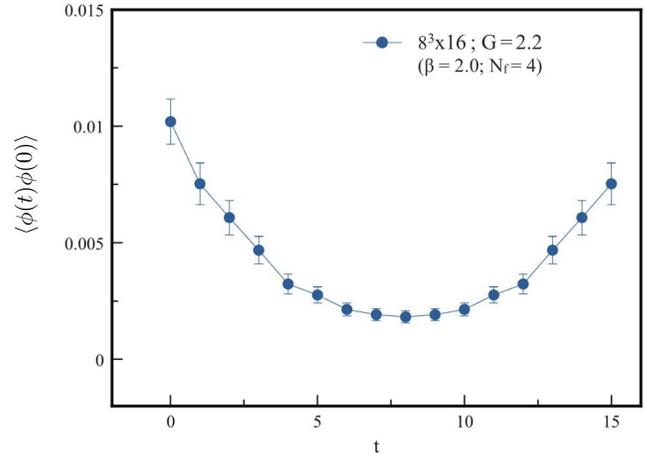}
\caption{Pion correlator for different $G$ at $\beta=2.2$ for $8^{3} \times 16$ lattice with $N_f=4$}
\label{oct16-sigma0-L816N4_beta2}
\end{center}
\end{figure}

In the continuum limit we nevertheless expect that the discrete lattice chiral symmetry will be enhanced to the full
continuous symmetry $SU_L(2)\times SU_R(2)$. In this case we expect
the auxiliary fields $\phi_i,\,i=1\ldots 3$ to behave as would be Goldstone bosons. Evidence in
favor of this is shown in
Figure~\ref{oct16-sigma0-L816N4_beta2}.
which shows a plot of $\langle\phi_1(t)\phi_1(0)\rangle$
for lattices $8^{3}\times 16$ at $\beta = 2.0$ in the strongly broken regime with four fermi coupling $G=2.2$.  We see
indeed that the auxiliary fields have developed dynamics and propagate as light quasi Goldstone bosons.

\section{Summary}

In this paper we have conducted numerical simulations
of the gauged NJL model for four flavors of Dirac fermion in the
fundamental representation of the $SU(2)$ gauge group.
We have employed a reduced staggered fermion discretization scheme which allows us
to maintain an exact subgroup of the continuum
chiral symmetries.

We have examined the model for a variety of values for lattices size, gauge coupling, and four fermi interaction strength. In the NJL limit $\beta\to\infty$
we find evidence for a continuous phase transition for $G\sim 1$ corresponding to
the expected spontaneous breaking of chiral symmetry. However, for gauge couplings that
generate a non-zero chiral condensate even for $G=0$ this transition or crossover appears
much sharper and  there is no evidence of critical fluctuations in the
chiral condensate.

Thus our results are consistent  with the idea that the second order phase transition which exists in the pure
NJL theory ($\beta=\infty$) survives at weak gauge coupling. However our results indicate that
any continuous transition ends if the gauge coupling becomes strong enough to cause confinement.
In this case we do however see evidence of additional dynamical mass generation
for sufficiently large four fermi coupling associated with
an observed rapid crossover in the chiral condensate and a possible first
order phase transition.
These results are consistent with the
numerical solution of an augmented ladder calculation \cite{Takeuchi:1989qa} reviewed in Section~\ref{sec:Continuum}.

The fact that we find the condensate non-zero and constant for strong
gauge coupling and $G< G_\text{cross}$ shows that the chiral symmetry of the theory is already broken as expected for $SU(2)$ with $N_f=4$ flavors.  This
breaking of chiral symmetry due to the gauge interactions is accompanied by the
generation of a non-zero fermion mass even for small four fermi coupling. Notice that
this type of scenario is actually true of top quark condensate models in which the
strong QCD interactions are already expected to break chiral symmetry
independent of a four fermion top quark operator.
The magnitude of
this residual fermion mass is {\it not} controlled by the four fermi coupling and cannot
to sent to zero by tuning the four fermi coupling - there can be no continuous phase
transition in the system as we increase the four fermi coupling - rather the condensate
becomes strongly enhanced for large $G$.

\begin{acknowledgments}
SMC and JH are supported in part by DOE grant
DE-FG02-85ER40237. SMC and JH would like to thank the Center for Particle Physics
Phenomenology $CP^3$ at the University of Southern Denmark
for support during the early phase of this work and
Francesco Sannino and Claudio Pica for important conversations.
The simulations were carried out using USQCD
resources at Fermilab and Jlab.
\end{acknowledgments}



\end{document}